# MODELINGTOOLKIT: A COMPOSABLE GRAPH TRANSFORMATION SYSTEM FOR EQUATION-BASED MODELING[*]

YINGBO MA[†], SHASHI GOWDA[‡], RANJAN ANANTHARAMAN[‡], CHRIS LAUGHMAN[§], VIRAL SHAH[¶], AND CHRIS RACKAUCKAS[‖]

**Abstract.** Getting good performance out of numerical equation solvers requires that the user has provided stable and efficient functions representing their model. However, users should not be trusted to write good code. In this manuscript we describe ModelingToolkit (MTK), a symbolic equation-based modeling system which allows for composable transformations to generate stable, efficient, and parallelized model implementations. MTK blurs the lines of traditional symbolic computing by acting directly on a user's numerical code. We show the ability to apply graph algorithms for automatically parallelizing and performing index reduction on code written for differential-algebraic equation (DAE) solvers, "fixing" the performance and stability of the model without requiring any changes on the user's part. We demonstrate how composable model transformations can be combined with automated data-driven surrogate generation techniques, allowing machine learning methods to generate accelerated approximate models within an acausal modeling framework. These reduced models are shown to outperform the Dymola Modelica compiler on an HVAC model by 590x at 3% error. Together, this demonstrates MTK as a system for bringing the latest research in graph transformations directly to modeling applications.

**1. Introduction.** Numerical solvers, such as those for differential equations and nonlinear systems, require that the modeler supply a function $f$ that appropriately defines their system. However, said modelers have a repeated history of writing bad code. This is not entirely the fault of users, since solver software places responsibility upon the user for the vast majority of stability and performance. Considering differential-algebraic equation (DAE) solvers as an example, high performance requires that the caller supply a function $f$ that is written in a highly performant way, parallelism requires that the caller supply an $f$ that is parallel, and stability requires that the caller supply an $f$ representing their model in an index-1 form [23, 11]. The main challenge is thus not the capabilities of the user, but rather the requirements placed upon the user by standard solver software.

There have been many attempts to abstract the user away from writing code for the solver to reduce the likelihood that poor model code influences solver performance. The most notable of these attempts are equation-based modeling systems, such as Modelica [18, 9, 31], which have steadily gained popularity by simplifying the modeling process [28]. In such systems a user describes the model at a high level by specifying differential equations and algebraic relationships. This allows a compiler implementation, such as Dymola [4], Modia.jl [5], or OpenModelica [7, 8], to generate high performance implementations for the underlying solvers such as DASSL [24] or SUNDIALS IDA [13]. While these numerical methods are also used in open source communities, modelers typically see better results in these systems due to the use of complex transformations before the application of numerical methods. These compilers perform a series of symbolic manipulations and graph algorithms, such as tearing of nonlinear systems [6] and methods for index reduction [22], to generate code that is more stable and faster than is normally achievable by hand. One significant downside of these declarative equation-based systems is that they are written in an environment that is separate from standard imperative languages (such as Python, MATLAB, C++, and Julia), making it difficult for users to customize the lowering process and help improve the code generation. This also means that users of these common programming languages must rewrite their models in a separate language to achieve better performance, increasing the number of man hours that are required in the modeling process.

---

[*]Submitted to the editors DATE.
 **Funding:** Anonymized for submission.
[†]Julia Computing and University of Maryland, Baltimore County
[‡]Massachusetts Institute of Technology,
[§]Mitsubishi Electronic Research Laboratory,
[¶]Julia Computing,
[‖]Massachusetts Institute of Technology and Pumas-AI

To counteract these issues, we present a new approach to automating the interface between modeler and solver with ModelingToolkit.jl (MTK), a Julia-based [3] modeling system which is directly embedded into its host language. We show how MTK removes barriers by demonstrating how symbolic index reduction can be performed on the user's original numerical code, generalizing ideas from techniques like automatic differentiation to the realm of automated code stability enhancements. This is coupled with automatic parallelism from extracted information about strongly connected components (SCC). We additionally demonstrate the use of automated approximation to show how the symbolic layer can be interfaced with numerical algorithms like those from machine learning to automatically replace sub-components of models with accelerated surrogates. These demonstrated capabilities establish MTK as a customizable modeling environment capable of automated code enhancement for modelers.

**2. MTK Overview.** In its simplest form, MTK is a symbolic equation-based modeling system written in Julia which obeys the semantics of its host language. For the purpose of this project, the authors built Symbolics.jl, a general-purpose Computer Algebra System (CAS) in the Julia programming language, as the core expression representation for MTK, which allows for efficient application of expression rewriting similar to scmutils [30]. MTK's symbolic specification of numerical problems is given via high level extension types defined by Symbolics.jl expressions. The modeling domain of MTK covers the span of:

- Ordinary differential(-algebraic) equations (`ODESystem`)
- Stochastic differential(-algebraic) equations (`SDESystem`)
- Partial differential equations (`PDESystem`)
- Nonlinear solve equations (`NonlinearSystem`)
- Optimization problems (`OptimizationSystem`)
- Continuous-Time Markov Chains (`JumpSystem`)
- Nonlinear Optimal Control (`ControlSystem`)

Further systems, like discrete-time Markov chains, are planned additions. Given that the symbolic analysis and discrete graph algorithms underlying the optimizations in each of these systems is similar, we will focus on the `ODESystem`in the remainder of this manuscript.

**2.1. Syntax and Generating Systems.** Expression building in MTK is done via overloading the operations of the `Num` type on standard Julia calls. While we will explore alternative uses of this feature in Section 3, its core purpose is to enable a natural syntax for symbolic specifications. For example, the following code builds and solves the classic Lorenz '63 system:

```
using ModelingToolkit, OrdinaryDiffEq
@parameters t σ ρ β
@variables x(t) y(t) z(t)
D = Differential(t)
eqs = [D(x) ~ σ*(y-x),D(y) ~ x*(ρ-z)-y,D(z) ~ x*y - β*z]
@named lorenz1 = ODESystem(eqs)
u0 = [x => 1.0,y => 0.0,z => 0.0]
p  = [σ => 10.0, ρ => 28.0, β => 8/3]
tspan = (0.0,100.0)
prob = ODEProblem(lorenz1,u0,tspan,p)
sol = solve(prob,Tsit5())
```

As an acausal modeling system, one can also arbitrarily compose system components and generate differential-algebraic equations by placing algebraic relations between the variables. The following extends the previous example to show how to connect a second instance of the Lorenz `ODESystem` to the first, with an algebraic variable defined implicitly by their connection:

```
@named lorenz2 = ODESystem(eqs)
@variables a; @parameters γ
connections = [0 ~ lorenz1.x + lorenz2.y + a*γ]
@named connected = ODESystem(connections,t,[a],[γ],systems=[lorenz1,lorenz2])
u0 = [lorenz1.x => 1.0,lorenz1.y => 0.0,lorenz1.z => 0.0,
      lorenz2.x => 0.0,lorenz2.y => 1.0,lorenz2.z => 0.0,
      a => 2.0]
p  = [lorenz1.σ => 10.0,lorenz1.ρ => 28.0,lorenz1.β => 8/3,
      lorenz2.σ => 10.0,lorenz2.ρ => 28.0,lorenz2.β => 8/3,
      γ => 2.0]
tspan = (0.0,100.0)
prob = ODEProblem(connected,u0,tspan,p)
sol = solve(prob,Rodas4())
```

**2.2. Stable Transformations of Systems.** The documentation delves further into the equation-based modeling system by showing how to generate large systems of differential-algebraic equations (DAEs) from acausal models. For example, a tutorial on circuit modeling showcases how individual circuit components, such as a resistor and capacitor, can be created separately and pieced together through connection equations to build an RC circuit. While this form of modeling can generate many redundant equations, a graph-based symbolic simplification function, `structural_simplify` (described further in Section 4), removes the numerically-unnecessary modeling abstractions to solve a single differential equation.

Functions of this sort are known as `transformations`. A transformation in MTK is a function which takes in a given system and returns a new system. Similar in spirit to composable transformations in the LLVM compiler [15, 19], we can effectively compose such transformations in any order by imposing a standard input and output representation on these transformations. For example, `dae_index_lowering` transforms a higher index DAE into an index-1 DAE via Pantelides algorithm [22] or its extensions [29], which can be composed with the singularity elimination (removal of $0 = 0$ relationships) of `alias_elimination` [20], `tearing` for reducing the size of nonlinear systems, and more. Similar passes are often incorporated into the compilation process of standard acausal modeling systems such as the Dymola compiler for Modelica [4] where the latter two are combined in MTK under the alias `structural_simplify`.

Given this open, stable, and composable compilation process, users can easily build new transformations which can be injected into the pipeline. For example, prior work has demonstrated that probabilistic robustness can be assessed via the trace of the Jacobian of an ODE [12]. One can easily perform such an analysis on an arbitrary DAE system by first running the `dae_index_lowering` before `structural_simplify`, after which `liouville_transform(sys)` can be used to generate the extra equation before solving. Because MTK is built on Symbolics.jl, all of the features of the CAS are available to the writer of the transformation, giving `liouville_transform` a simple 7 line implementation with the core being to append the term `D(trJ)~trJ*-tr(calculate_jacobian(sys))`.

MTK can thus be seen to provide a Modelica-like equation-based modeling system which follows Julia's syntax but allows for composing user-made transformations. In the following sections we showcase of the emergent features of this combination.

**3. Index Reduction of Numerical Code.** While index reduction of DAEs is commonly performed in equation-based modeling systems to enhance the stability of the DAE solving process, such symbolic enhancements can be applied directly to a user's code using abstract interpretation. Recall that the semantics of a Symbolics.jl expression directly match those of Julia itself. This means that for some symbolic variable `x` and some arbitrary Julia numerical program `f`, by operator overloading a symbolic representation of a user program can be generated simply by calling `f(x)`[1]. Given this abstract interpretation capability, MTK allows ODE definition code written for the DifferentialEquations.jl solvers [27] to be automatically converted to MTK's symbolic representation via the `modelingtoolkitize` function.

To illustrate the utility of this functionality, let's investigate the two-dimensional pendulum without small angle approximations. The system of equations is simple: it is given by 5 state variables, $x$, $y$, the velocities $v_x$ and $v_y$, and $T$, with parameters $g$ for the acceleration due to gravity and the fixed length of the pendulum $L$. This system is as follows:

$$x' = v_x \tag{3.1}$$
$$v_x' = Tx \tag{3.2}$$
$$y' = v_y \tag{3.3}$$
$$v_y' = Ty - g \tag{3.4}$$
$$0 = x^2 + y^2 - L^2 \tag{3.5}$$

A user of DifferentialEquations.jl may thus wish to solve the equation by writing code for the numerical solvers in form:

```
function pendulum!(du, u, p, t)
    x, dx, y, dy, T = u
    g, L = p
    du[1] = dx; du[2] = T*x
    du[3] = dy; du[4] = T*y - g
    du[5] = x^2 + y^2 - L^2
```

---

[1] Note that some restrictions on the Julia program are currently required, mainly the program must be "quasi-static", as in a program representation must not be dependent on the input values. For example, a while loop can be handled only if the number of steps for termination can be computed without knowing the numerical value of x.

```
end
pendulum_fun! = ODEFunction(pendulum!, mass_matrix=Diagonal([1,1,1,1,0]))
u0 = [1.0, 0, 0, 0, 0]; p = [9.8, 1]; tspan = (0, 10.0)
pendulum_prob = ODEProblem(pendulum_fun!, u0, tspan, p)
```

This type of equation is known as an index-3 DAE because the Jacobian of the algebraic equation with respect to $L$ is singular. Since most DAE solvers require an index-1 DAE, a user may easily and inadvertently write such an "unsolvable" equation, leading to a non-convergent solving process and leaving the user to think the issue is a "bad solver". For example, if the user attempts to solve this with the `Rodas4` method for DAEs, the solver almost immediately exits due to claims of instability.

Because the underlying problem is structural, we can use the `modelingtoolkitize` function to transform the user's equations to MTK symbolic representations, and subsequently use `dae_index_lowering` to generate an index-1 form which is numerically solvable. The algebraic equation can then be replaced via a rootfind inside the right hand side system (done as part of the `ODAEProblem` construction) to arrive at a simple non-stiff ODE. These transformations are performed via:

```
traced_sys = modelingtoolkitize(pendulum_prob)
pendulum_sys = structural_simplify(dae_index_lowering(traced_sys))
prob = ODAEProblem(pendulum_sys, Pair[], tspan)
sol = solve(prob, Tsit5())
```

The method automatically transforms the user's code to the form:

$$
\begin{align}
x' &= v_x \tag{3.6}\\
v_x' &= xT \tag{3.7}\\
y' &= v_y \tag{3.8}\\
v_y' &= yT - g \tag{3.9}\\
0 &= 2\left(v_x^2 + v_y^2 + y(yT - g) + Tx^2\right) \tag{3.10}
\end{align}
$$

which is mathematically equivalent but numerically more stable. The graph algorithm automatically figures out that only the fifth equation needs to be changed, and it needs to be differentiated twice for the numerical solution to be stable. This variant of the model is successfully solved with an explicit Runge-Kutta method for non-stiff ODEs [32], as shown in Figure 1. Such automated transformation features allows users of the numerical environment to make use of symbolic graph algorithms like Pantelides algorithm to nearly invisibly achieve better numerical stability. This abstract interpretation process thus enables the application of such discrete algorithms by the users of numerical functionality of standard programming languages.

**4. Extracting Parallelism from Structural Simplification.** The previous section demonstrated that the stable simulation may require a different model form from that which is supplied by the user. This is of particular concern when attempting to parallelize a model, as the user must parallelize the same function to be replaced. Unfortunately, this user will not generally know what final equations will result after the transformations. Optimal parallelism with the numerical solution of differential equation therefore must be achieved via application of an automated analysis which can be composed with the structural transformations of the model code.

Our system can auto-parallelize on two sets of computations: the right-hand side of the differential equations, and the computation of the torn algebraic variables. The former parallelization is straightforward, since the right-hand side expressions are embarrassingly parallel; in comparison, the latter optimization is more challenging and can be achieved by analyzing the graph structure created in the tearing process. During the `structural_simplify` transformation, the tearing process [20] finds the strongly connected components graph. This information describes which components are either independent or only connected via differentiated variables. Analysis of this graph enables a directed acyclic graph of these dependencies between components to be constructed, and the groups of required non-linear solver calls to compute the algebraic variables are spawned in concurrent threads.

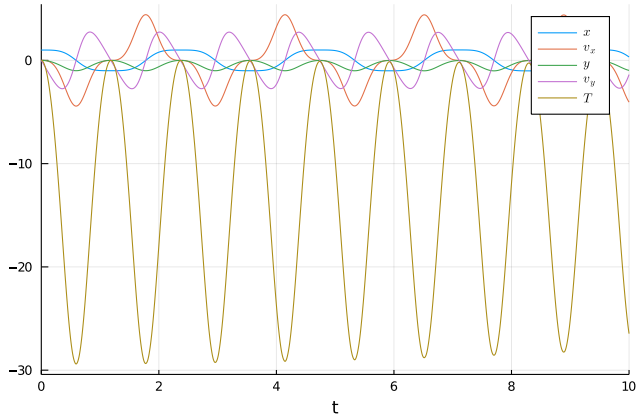

Fig. 1: **Index-3 DAE model of a pendulum solved by an explicit Runge-Kutta method**. Shown is the pendulum model successfully solved by the Tsit5 [32] 5th order Runge-Kutta method after performing index reduction via the Pantelides algorithm and reducing the nonlinear system. We emphasize that this is performed purely on the user's code for the numerical solver, not requiring a separate equation-based modeling language, and the integration fails without the transformation.

In Figure 2 we show the incidence matrix of 50 independent RC circuits whose resistance is a non-linear function of the ambient temperature, and that of a model of an HVAC system [17, 16, 26]. A non-zero $(i, j)$-th entry means $j$-th variable appears in $i$-th equation. Each connected component in the subfigures in the second and third column imply a single call to solver to obtain the values of the algebraic variables in that component. Our system is able to detect independence between components and spawn these calls in parallel. Note that we only show algebraic equations and algebraic variables in the Block Lower Triangular (BLT)-sorted spy plots, because only the algebraic part of the system has a non-trivial dependency structure that we need to analyze for parallel executions.

These results demonstrate why composing the transformations is important: by performing the graph simplifications before parallelization, we only end need to numerically solve an order of magnitude fewer equations than the user's model originally specified, while exposing more parallelism.

**5. Data-Driven Model Order Reduction Transformations.** While the aforementioned transformations were exact, this is not a requirement of the composable transformation rules. One immediate consequence is that methods for model approximation, such as model order reduction or the creation of data-driven surrogates, can be automated in this modeling environment as transformations. Such surrogates can then be embedded in larger systems defined within MTK itself for accelerating design, controls, and optimization. This is not possible in prior equation-based modeling environments because such approximation require domain-specific user input into the compilation pass structure in order to achieve a desired accuracy.

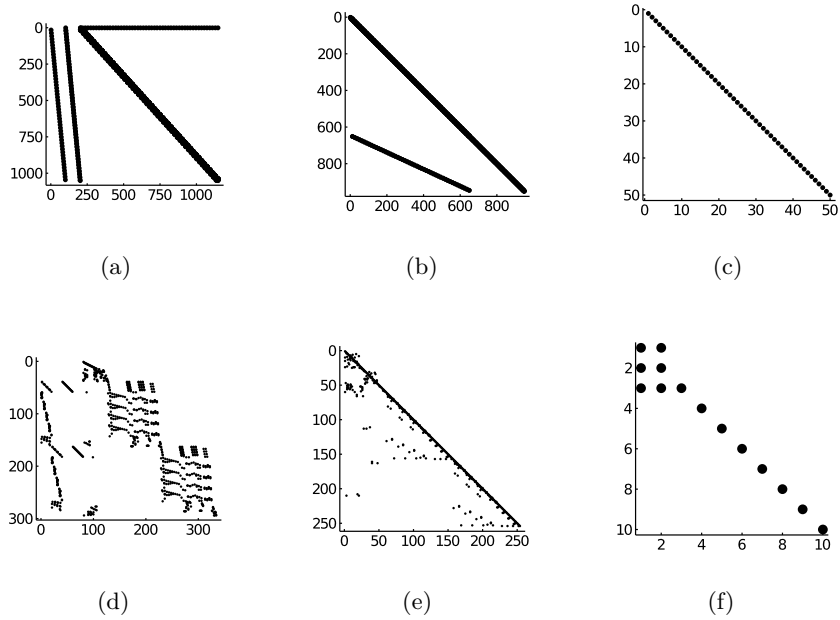

Fig. 2: Incidence matrices of two systems at various stages of the lowering process. Block diagonal components are independent and can be solved in parallel. (a) system of 50 independent RC circuits with resistances. (1051 × 1152) (b) BLT reordering of the former (950 × 950) (c) After tearing. (50 × 50) (d) Original incidence matrix of a HVAC system (294 × 334) (e) After BLT reordering (254 × 254) (f) After tearing (10 × 10).

To demonstrate the applicability of this process, we created a transformation `surrogatize` which takes in the user's ODE and a parameter space $P$ on which to train a continuous-time echo state network, a surrogate which has been shown to be robust to stiffness [2]. This transformation samples $P$ to train a continuous surrogate equation of the form:

$$r' = f(Ar + W_{fb}x(p^*, t)) \tag{5.1}$$
$$x(t) = g(W_{out}(p)r(t)) \tag{5.2}$$

where $p^* \in P$ is a fixed parameter, $A$ and $W_{fb}$ are fixed matrices, and $W_{out}(p)$ is a learned projection function determined by the training process.

To demonstrate the composability, we performed this approximate transformation on an 8,000 equation model of a heating cycle within an HVAC system [17, 16, 26]. The system was trained on 8 input parameters and 16 outputs. As a large DAE system, the training process for this model requires transformations to be stabilized for the numerical solvers, which was achieved by performing `dae_index_lowering` before `structural_simplify` for singularity elimination and tearing. On this reformulation of the model, we run the `surrogatize` transformation to generate an approximate surrogate model. The accuracy of the surrogatized system with reservoir size (length of $r$) at 1000 are shown in Figure 3 and Table 4. These results show that the automated reduction transformation found a new `ODESystem` definition with simulation results within 3% error over the chosen input parameter space. The generated surrogate allows for solving the system in 0.06 seconds, approximately 97x faster than the original 5.8 seconds and around 590x faster than the equivalent model solved in Dymola in 35.3 seconds with the modified DASSL solver. Given that the output of `surrogatize` is simply another `ODESystem`, it can be used as a subcomponent of a larger model, such as is shown in Section 2.1. In this way, a building model can replace its high fidelity HVAC model with an order of magnitude faster machine learned surrogate by using one function call.

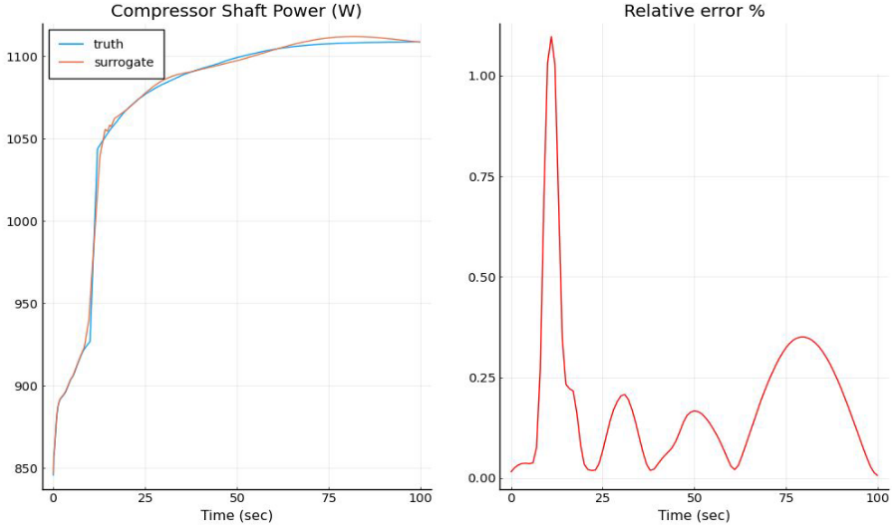

Fig. 3: **Acurracy of HVAC surrogate generated by MTK on a given output**: The blue line represents the ground truth from the high fidelity simulation and the red line shows the prediction from the surrogate. Values are taken from a test parameter not included in the training dataset.

| Accuracy of surrogate on quantities of interest | | | |
| --- | --- | --- | --- |
| Output | Error (%) | Output | Error (%) |
| Compressor inlet pressure | 0.788 | Condenser heat flow rate | 0.792 |
| Compressor outlet pressure | 0.732 | Evaporator heat flow rate | 2.28 |
| LEV inlet pressure | 0.811 | Coefficient of Performance | 0.281 |
| LEV outlet pressure | 0.378 | Condenser outlet air temperature | 0.009 |
| LEV outlet specific enthalpy | 2.65 | Evaporator outlet air temperature | 0.002 |
| Compressor refrigerant mass flow rate | 2.21 | Compressor power consumption | 1.097 |
| LEV refrigerant mass flow rate | 1.839 | Compressor inlet specific enthalpy | 1.858 |
| Evaporator superheat temperature | 2.33 | Compressor outlet specific enthalpy | 1.978 |

Fig. 4: **Accuracy of surrogate on all output quantities:** The CTESN surrogate was tested on 16 output quantities of interest from the HVAC system, demonstrating within 3% accuracy.

**6. Conclusions.** While discrete graph-based algorithms have been added before to equation-based modeling environments to improve stability and performance, MTK is the first to embed such a modeling environment directly into a high performance host language and allow for composable transformations to be constructed using a CAS. In a sense, this extends the abstract interpretation techniques of automatic differentiation to directly perform arbitrary symbolic transformations on a user's code for alternative mathematical purposes. We demonstrated three use cases that arise:

1. Automating the DAE index reduction.
2. Parallelism of equations after nonlinear tearing and singularity elimination.
3. Automated data-driven model order reduction in a simulation environment.

This software environment opens many avenues for researchers in discrete and symbolic algorithms to directly apply their work to a wide variety of models. As an open source software, anyone can add new passes to compose as part of the system. We see this as especially helpful for accelerating the research of data-driven and machine learning approaches for model order reduction, as this system can help automate the process and begin testing against alternative techniques within full-scale applications.

There are many more transformation passes which have not been explored. To end, we list a few ideas the authors wish to work with the open source community to implement:

- Automatic restructuring of equations for improved floating point accuracy with algorithms such as Herbie [21].
- Allowing delays in acausal modeling via extensions to the Pantelides algorithm for delay differential equations [1].
- Automating the application of model order reduction techniques such as Proper Orthogonal Decomposition (POD) [14] and Lift & Learn [25].
- Automating model analyses, such as structural identifiability of parameters [34, 10, 33], on the structurally-simplified index-reduced resulting equations and directly on a user's numerical code.

The authors know of no other equation-based acausal modeling systems which are open source and easily extendable with such composable transformation and analysis passes, and thus anticipate that the MTK environment will play an important role as a bridge for researchers interested in testing symbolic and discrete algorithms in this domain.

**Acknowledgments.** This material is based upon work supported by the National Science Foundation under grant no. OAC-1835443, grant no. SII-2029670, grant no. ECCS-2029670, grant no. OAC-2103804, and grant no. PHY-2021825. We also gratefully acknowledge the U.S. Agency for International Development through Penn State for grant no. S002283-USAID. The information, data, or work presented herein was funded in part by the Advanced Research Projects Agency-Energy (ARPA-E), U.S. Department of Energy, under Award Number DE-AR0001211 and DE-AR0001222. We also gratefully acknowledge the U.S. Agency for International Development through Penn State for grant no. S002283-USAID. The views and opinions of authors expressed herein do not necessarily state or reflect those of the United States Government or any agency thereof. This material was supported by The Research Council of Norway and Equinor ASA through Research Council project "308817 - Digital wells for optimal production and drainage". Research was sponsored by the United States Air Force Research Laboratory and the United States Air Force Artificial Intelligence Accelerator and was accomplished under Cooperative Agreement Number FA8750-19-2-1000. We additionally thank Simon Frost of Microsoft Research Studies in Pandemic Preparedness for helping fund this work. The views and opin-